\documentclass[twocolumn,showpacs,superscriptaddress,preprintnumbers,amsmath,amssymb,prb]{revtex4}
\usepackage{natbib}

\usepackage{graphicx}
\usepackage{dcolumn}
\usepackage{bm}

\begin{document}


\title{GaN:$\delta$-Mg grown by MOVPE: structural properties and their\\effect
on the electronic and optical behaviour}

\author{Tian Li}
\email{Tian.Li@jku.at}
\affiliation{Institut f\"{u}r Halbleiter- und Festk\"{o}rperphysik, Johannes Kepler University\\ Altenbergerstr. 69, A-4040 Linz, Austria}

\author{Clemens Simbrunner}
\affiliation{Institut f\"{u}r Halbleiter- und Festk\"{o}rperphysik, Johannes Kepler University\\ Altenbergerstr. 69, A-4040 Linz, Austria}

\author{Matthias Wegscheider}
\affiliation{Institut f\"{u}r Halbleiter- und Festk\"{o}rperphysik, Johannes Kepler University\\ Altenbergerstr. 69, A-4040 Linz, Austria}

\author{Andrea Navarro-Quezada}
\affiliation{Institut f\"{u}r Halbleiter- und Festk\"{o}rperphysik, Johannes Kepler University\\ Altenbergerstr. 69, A-4040 Linz, Austria}

\author{Martin Quast}
\affiliation{Institut f\"{u}r Halbleiter- und Festk\"{o}rperphysik, Johannes Kepler University\\ Altenbergerstr. 69, A-4040 Linz, Austria}

\author{Klaus Schmidegg}
\affiliation{Institut f\"{u}r Halbleiter- und Festk\"{o}rperphysik, Johannes Kepler University\\ Altenbergerstr. 69, A-4040 Linz, Austria}

\author{Alberta Bonanni}
\affiliation{Institut f\"{u}r Halbleiter- und Festk\"{o}rperphysik, Johannes Kepler University\\ Altenbergerstr. 69, A-4040 Linz, Austria}

\begin{abstract}

The effect of Mg $\delta$-doping on the structural, electrical and optical properties of GaN grown $\textsl{via}$ metalorganic vapor phase epitaxy has been studied using transmission electron microscopy, secondary ion mass spectroscopy, atomic force microscopy, x-ray diffraction, Hall effect measurements and photoluminescence. For an average Mg concentration above 2.14 $\times$ 10$^{19}$ cm$^{-3}$, phase segregation occurs, as indicated by the presence of Mg-rich pyramidal inversion domains in the layers. We show that $\delta$-doping promotes, in comparison to Mg continuous doping, the suppression of extended defects on the samples surface and improves significantly the morphology of the epilayers. Conversely, we can not confirm the reduction in the threading dislocation density - as a result of $\delta$-doping - reported by other authors. In the phase separation regime, the hole concentration is reduced with increasing Mg concentration, due to self-compensation mechanisms. Below the solubility limit of Mg into GaN at our growth conditions, potential fluctuations result in a red-shift of the emission energy of the free-to-bound transition.

\end{abstract}

\maketitle

\section{\label{sec:level1} Introduction}

Group III nitrides represent one of the most important family of semiconductors for optoelectronic devices because of their direct band gap, which spans the range including the whole of the visible spectrum and extending well into the ultraviolet (UV) and infra-red (IR) regions.\cite{TOM} Furthermore, nitride diluted magnetic semiconductors (DMS) doped with transition metals or rare earths and containing a sufficient concentration of acceptors, are expected to show ferromagnetic behaviour above room temperature (RT) and to be, therefore, promising materials for spintronics applications.\cite{Dietl} However, due to the large $n$-type background in the as-grown layers, high $p$-type doping remains one of the greatest challenges in the fabrication of nitride-based high performance devices. Mg is recognized as the most efficient source of acceptors for nitrides and hindrances originating from the passivation effect of Mg-H complexes\cite{naka1} have been successfully relieved by low-energy electron beam irradiation (LEEBI)\cite{amano} and thermal annealing.\cite{naka} Still unsolved is the problem of the self-compensation,\cite{NVM,KSO} which limits the free hole concentration to 10$^{18}$ cm$^{-3}$ for a Mg content in the 10$^{19}$ cm$^{-3}$.\cite{OBK}
\indent $\delta$-doping has been proposed as a promising way to enhance the Mg incorporation into a GaN matrix and to promote the free carrier concentration in the layers.\cite{NST,SKK} Moreover, it has been reported to reduce the density of threading dislocations in the highly faulted nitride layers and, therefore, to improve the crystalline structure of the host semiconductor.\cite{CPC}

\indent Recent works have reported on the enhanced electrical efficiency of GaN:$\delta$-Mg,\cite{NKJ,WLN} when compared to homogeneously doped structures, with improved surface morphology attributed to the reduction of threading dislocations upon $\delta$-doping. Other authors focus on the defects in GaN:$\delta$-Mg, underlining the analogy with faults in GaN:Mg. \cite{web1,web2,web3} Still controversial is the issue whether the insertion of Mg into GaN in a $\delta$-fashion represents an effective alternative to the homogeneous doping. In order to shed new light on this matter, we have already
reported the actual periodic distribution of Mg ions in heavily doped GaN:$\delta$-Mg layers.\cite{SWQ} In this work, we determine the solubility limit of $\delta$-Mg in GaN at the given growth conditions and we provide a quantitative analysis of the distribution of Mg-rich defects originating from phase separation. By studying cross-sectional specimens, we get insights into the improved surface morphology of GaN:$\delta$-Mg with respect to that of GaN:Mg. Furthermore, we discuss the influence of Mg $\delta$-doping on electrical and optical properties.

\begin{table*}
\caption{\label{tab:table1} Growth parameters and structural, electrical and optical data for the investigated GaN:$\delta$-Mg samples}
\begin{ruledtabular}
\begin{tabular}{cccccccccc}
&Cp$_2$Mg&Time&SIMS&XRD&Hole conc.&N$_a$&PL\\
ID&(sccm)&(s)&10$^{19}$cm$^{-3}$&FWHM&10$^{17}$cm$^{-3}$&10$^{19}$cm$^{-3}$&color\\\hline

 No.495&300&25&1.87&253&2.53&7.31&VL\\
 No.499&350&30&1.99&255&2.75&8.77&VL\\
 No.503&350&30&2.07&242&1.74&11.4&VL\\
 No.497&350&25&2.14&249&3.38& -- &VL\\
 No.493&400&25&2.71&250&2.29&9.75&BL\\
 No.496&300&35&2.50&287&2.41&10.2&BL\\
 No.501&350&30&2.43&245&2.71&10.2&BL\\
 No.500&400&30&2.75&286&2.48&12.1&BL\\
 No.498&350&35&3.14&286&2.01&12.4&BL\\
 No.494&400&35&3.68&296&1.59&14.1&BL\\
\end{tabular}
\end{ruledtabular}
\end{table*}

\section{Experiment}

The Mg-$\delta$-doped GaN epilayers have been fabricated on $\textit{c}$-plane sapphire substrates in an AIXTRON 200RF-S horizontal flow metalorganic vapour phase epitaxy (MOVPE) reactor. Trimethylgallium(TMGa), biscyclopentadienyl magnesium (Cp$_2$Mg), and ammonia (NH$_3$) were used as Ga, Mg and N sources, respectively. \cite{SWQ} Upon deposition of a 1$\mu$m GaN buffer layer at 1020 $^\circ$C, Mg-$\delta$-doped GaN was grown at a substrate temperature of 950 $^{\circ}$C by alternatively closing and opening the TMGa and the Cp$_2$Mg sources. The superlattice structures consist of 66 periods of alternate Mg-$\delta$-doped layers and GaN spacer layers, both with a nominal thickness of 13.22 nm. All samples have been annealed $\textit{in-situ}$ under nitrogen at 780~$^\circ$C for 15 minutes. Growth parameters including the Cp$_2$Mg flow rate and the exposure time of the surface to Mg, have been controlled in order to optimize the Mg-incorporation.

In Table \ref{tab:table1}, the structural, electrical, chemical and optical parameters for Mg $\delta$-doped GaN epilayers grown respectively with three different Cp$_2$Mg flow rates [300 standard cubic centimeters per minute (sccm), 350 sccm and 400 sccm] and varying Mg deposition time are reported.
The growth process has been monitored by means of \textit{in-situ} laser reflectometry (LR) \cite{SIJAP} and x-ray diffraction.\cite{SIJCG} The Mg concentration in the layers was evaluated \textit{via} secondary ion mass spectrometry (SIMS) by employing sputtering with O$_{2}$ and Cs$^{+}$ for the detection of Mg and hydrogen, respectively. Photoluminescence (PL) experiments were carried out with the 325 nm line of a He-Cd laser as excitation source. Hall effect was measured in  Van der Pauw geometry in the temperature range from 150 to 440 K. High-resolution transmission electron microscopy (HRTEM) studies have been performed on cross-sectional samples prepared by standard mechanical polishing followed by Ar ion milling at 4 KV for about 1 hour. The TEM images were obtained from a JEOL 2011 Fast TEM microscope operated at 200 kV and equipped with a Gatan CCD camera.

\begin{figure}
\includegraphics{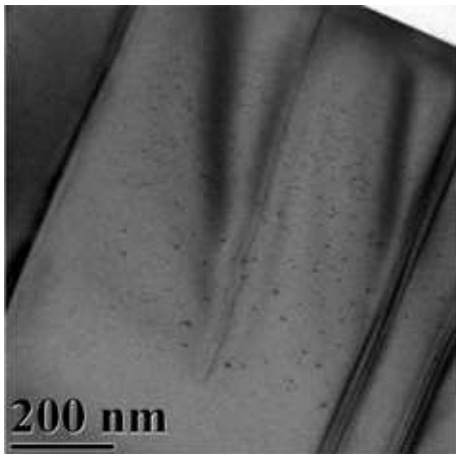}
\caption{\label{fig:delta} Micrograph of a Mg-$\delta$-doped GaN layer (No.496), showing a non-homogeneous distribution of PIDs. Moving fom the sample surface to the nominally undoped GaN buffer, the size of PIDs increases and their density diminishes.}
\end{figure}
\section{results and discussion}

\begin{figure}
\includegraphics{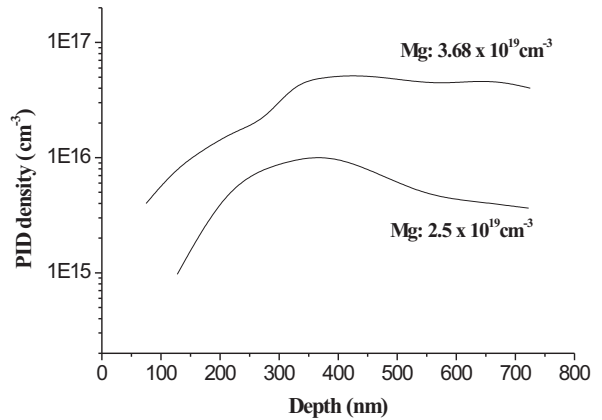}
\caption{\label{fig:density} PID density \textit{vs} depth from the surface for two GaN:$\delta$-Mg samples with respectively high (No.494) and low (No.496) Mg concentration.}
\end{figure}
\subsection{\label{sec:level2}Mg distribution in GaN:$\delta$-Mg}

SIMS analysis on GaN:$\delta$-Mg epilayers with an average total Mg concentration of 3.68 $\times$ 10$^{19}$ cm$^{-3}$ reported in one of our previous works \cite{SWQ} revealed indeed, as an effect of the incorporation in a $\delta$-fashion, a periodic distribution of the Mg ions. Moreover, our TEM studies on the same structures did not show neither compositional ordering in diffraction patterns nor periodic planar defects in HRTEM, implying that the Mg dopants are well-incorporated into the lattice sites
during the $\delta$-doping process. The employed growth conditions ensure, in the growth reactor, a nitrogen rich environment that promotes the Ga-polarity (N-termination) of the GaN spacer layers, as proven by the structural analysis of the Mg-rich defects. These growth conditions generate a large number of Ga vacancies suitable for the Mg ions incorporation. It has been reported that, in contrast, in Mg-doped GaN of N-polarity (Ga-terminated), periodic planar defects are formed by Mg accumulation.\cite{web5}

\subsubsection{Above the solubility limit of Mg into GaN}
For an average Mg concentration higher than 2.14 $\times$ 10$^{19}$ cm$^{-3}$, triangular defects are observed in cross-sectional TEM images. Similarly shaped faults have been reported for Mg continuously doped GaN grown by MOVPE,\cite{VBB,BVB} molecular beam epitaxy (MBE)\cite{RFS,RNP} and high pressure/high temperature techniques.\cite{web5,VBD,web4} These triangular defects have been identified as Mg-rich pyramidal inversion domains (PIDs), resulting from phase separation. As we will show later, HRTEM studies confirm that triangular defects in GaN:$\delta$-Mg are the profile of PIDs with the same structure as those found in GaN:Mg. From the above observations, we infer that 2.14 $\times$ 10$^{19}$ cm$^{-3}$ is the critical value for phase separation in GaN:$\delta$-Mg at our growth conditions.

\begin{figure}
\includegraphics{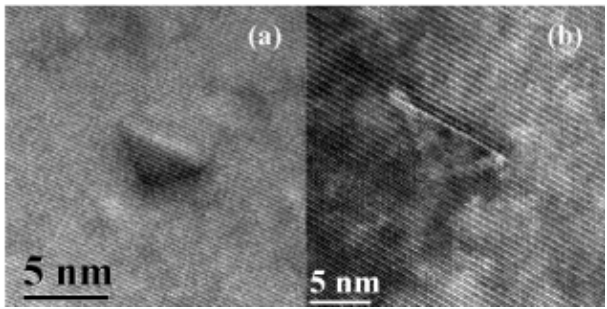}
\caption{\label{fig:PID} GaN:$\delta$-Mg: high resolution TEM images of PIDs on (a) the [10$\overline{1}0$ zone axis] and (b) in the [11$\overline{2}0$] zone axis.}
\end{figure}

In Fig.~\ref{fig:delta} the low-resolution TEM image of a GaN:$\delta$-Mg sample above the Mg solubility limit lets identify the inhomogeneous distribution of PIDs in the layer. A statistical evaluation of the inversion defects density in different samples shows that the average number of PIDs systematically increases with the Mg concentration. In Fig.~\ref{fig:density} the PIDs density for two  GaN:$\delta$-Mg epilayers with different Mg content, is reported as a function of the depth from the sample surface and it is found to be minimal in proximity of the surface, to increase to a saturation point with the depth, and to slightly decrease when approaching the interface with the nominally undoped GaN buffer layer.
We explain the observed variation of PID density in each layer by considering the way Mg diffuses into the GaN matrix during the growth process and assuming that the diffusion of the Mg ions into GaN obeys the Fick's first diffusion law, namely:

\begin{subequations}
\label{eq:fick}
\begin{equation}
\frac{\partial C_{Mg}}{\partial t} = -div \vec{J}_{Mg} ,\label{subeq:1}
\end{equation}
\begin{eqnarray}
\vec{J}_{Mg} = -D \cdot \nabla C_{Mg}.\label{subeq:2}
\end{eqnarray}
\end{subequations}

where $\vec{J}_{Mg}$ denotes the diffusion flux , $C_{Mg}$ the concentration on Mg ions and $D$ the diffusion coefficient. Eq.~\ref{eq:fick} implies that the diffusion flux of the Mg dopant is proportional to the gradient of its time-dependent concentration. The diffusion length $L_d$ can be described by the Einstein-Smoluchowski approximation:

\begin{equation}
L_{d}=\sqrt{Dt},\label{eq:ES}
\end{equation}

where $t$ is the time for diffusion. During the Mg $\delta$-doping process, the time suitable for diffusion of the Mg ions reduces monotonically with the proceeding of the growth. The $\delta$-Mg layers deposited in subsequent stages tend to be increasingly localized, since the decreasing time for diffusion reduces the diffusion length , this effect explaining the periodical distribution of Mg in the vicinity of the sample surface. Moreover, since the Mg dopants are rather localized, the surface of GaN:$\delta$-Mg is virtually free of extended defects, leading to an improved surface morphology. On the other hand, close to the interface with the nominally undoped GaN buffer, due to the longer diffusion time of the first $\delta$-layers, more Mg ions tend to aggregate to form larger PIDs with, consequently, a reduced density.

\begin{figure}
    \centering
        \includegraphics{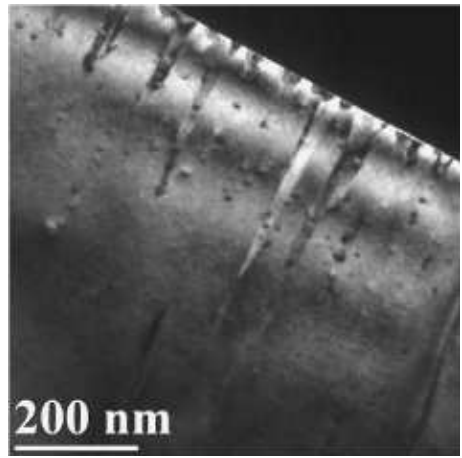}
    \caption{Distribution of PIDs in a Mg-continously-doped GaN. The Mg ions tend to segregate to the surface, where they aggregate in inversion domains.}
    \label{fig:bulk}
\end{figure}

HRTEM studies show that the microstructure of pyramidal defects in GaN:$\delta$-Mg is the same as the one of those observed in GaN:Mg. Fig.~\ref{fig:PID}(a) and ~\ref{fig:PID}(b) report the high resolution images of PIDs in a cross-sectional GaN:$\delta$-Mg specimen viewed along the [10$\overline{1}$0] and the [11$\overline{2}$0] zone axis, repectively. The angles between the basal plane and the faceted planes are 47$\pm1$$^{\circ}$ and 43$\pm1$$^{\circ}$ for the projections along the [10$\overline{1}$0] and [11$\overline{2}$0] direction, respectively.
Therefore, the PIDs have a base on the (0001) plane and six sidewalls on the $\left\{11\overline{2}3\right\}$ planes, exactly the same as those reported for the Mg-continuously-doped GaN.
In dark-field images (not shown here) at the two beam conditions, the pyramidal structures contrast reverses for g = 0002 and g = 000$\overline{2}$, confirming their nature of inversion domains.
The tip of the PIDs points toward the substrate, implying that the matrix has Ga-polarity.\cite{VBB} The average size of PIDs - i.e. the basal plane diameter - in our GaN:$\delta$-Mg layers is in the range 5-10 nm, smaller than the 5-20 nm reported for PIDs in GaN:Mg.\cite{PSD}

\subsubsection{Below the solubility limit of Mg into GaN}
For an average Mg concentration below 2.14 $\times$ 10$^{19}$ cm$^{-3}$, we could not detect phase separation. Furthermore, in this regime, SIMS measurements did not show a modulated Mg profile of the doped layers. Although, we can not rule out that in this case the local fluctuation in Mg composition lies below the detection limit of SIMS.

\subsection{PIDs distribution: GaN:$\delta$-Mg $\textit{vs}$ GaN:Mg}
For comparison purposes, GaN:Mg samples have been fabricated at the same growth conditions as the GaN:$\delta$-Mg structures above solubility limit, except that Mg has been continuously incorporated into the GaN matrix. The large number of PIDs and inversion domains visible in the low-resolution TEM of Fig.~\ref{fig:bulk} confirms that also in this GaN:Mg sample phase separation occurs, although here the distribution of PIDs is quite different from that in GaN:$\delta$-Mg: here both size and density of the inversion domains increase in the region close to the sample surface. Therefore - in contrast to the case of GaN:$\delta$-Mg - upon phase separation, in GaN:Mg the Mg ions tend to segregate to the surface, similar tendency being reported for MBE-grown GaN:Mg.\cite{LXZ,GBC}

A systematic atomic force microscopy (AFM) study of the surface morphology of both GaN:$\delta$-Mg and GaN:Mg samples allows to notice a trend in the root-mean-square (RMS) roughness as a function of the fashion of Mg incorporation and of the nominal Mg content. The GaN:$\delta$-Mg samples generally show a smooth-terraced surface virtually free of extended defects and with an RMS of 1~-~6~nm, linearly increasing with the Mg content. Most of the terraces are pinned by dips, associated to the intersection of screw or mixed threading dislocations with the free surface. \cite{YCO}
In contrast, the GaN:Mg surface presents a higher surface roughness, estimated in some 10~nm, likely to be correlated with the presence of inversion domains in proximity of the sample surface. This could be the main reason for the improved surface morphology of GaN:$\delta$-Mg. We would like to point out that our observations are in contrast with Ref. \onlinecite{WLN} where the formation of PIDs in GaN:$\delta$-Mg is ruled out, based on  surface measurements solely. The authors of both the Ref. \onlinecite{NKJ} and Ref. \onlinecite{WLN} argue that the reduction of threading dislocations, claimed to result from $\delta$-Mg doping, contributes decisively to the improved morphology. In the following section, we will discuss our observations of the influence of Mg-$\delta$-doping on the threading dislocations (TDs) in the GaN matrix.

\subsection{\label{sec:level3} Threading dislocations in GaN:$\delta$-Mg}
The epitaxial growth mode dramatically affects the interplay between TDs in the matrix and the Mg ions. LR data recorded $\textit{in-situ}$ indicate a two-dimensional (2D) growth mode for all investigated $\delta$-doped layers,\cite{SWQ} as confirmed by AFM measurements. The deposition of Mg in a $\delta$-fashion reduces periodically the reflectivity of the surface and this attenuation is either related to the different optical response of the incorporated Mg adatoms, or to the reduction of GaN layer thickness by etching effects.\cite{SWQ} The growth of the GaN spacers restores the 2D growth right after the Cp$_2$Mg precursor source is switched off. No nucleation process occurs during $\delta$-doping.
In the \textit{ex-situ} XRD measurement, the full-width-at-half-maximum (FWHM) of the rocking curves of the (0002) reflection range from 245 to 296 arcsec over the whole series of investigated samples. Therefore, the overall fluctuation of the FWHM is estimated in only one sixth, whereas the average Mg concentration doubles its value. This implies that the screw dislocation density, proportional to the square of the FWHM of the $\omega$-scans,\cite{MHB} does not change considerably with varying the Mg content in the GaN:$\delta$-Mg layers.
\begin{figure}
\includegraphics{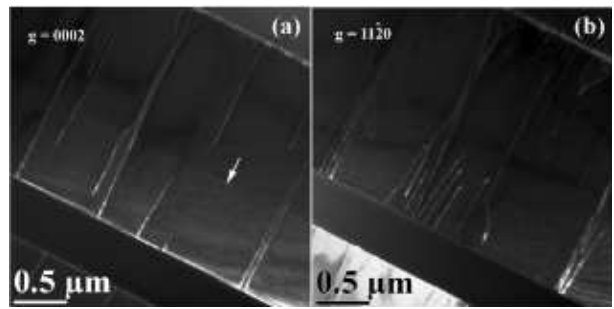}
\caption{\label{fig:dislocation} Weak-beam dark field  images of a GaN:$\delta$-Mg sample taken close to the [10$\overline{1}$0] zone axis with (a) g = 0002 (2g/5g) and (b) g = 11$\overline{2}$0 (1g/3g), respectively.}
\end{figure}

\begin{figure}
    \centering
        \includegraphics{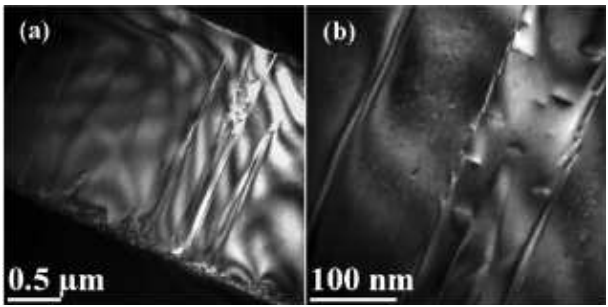}
    \caption{(a) TEM micrograph of the region surrounding two neighbour threading dislocations, where a pit associated to the two TD can be resolved at the sample surface. (b) High magnification image showing large PIDs in between the two TD, with the dislocations generating a bamboo-like void contrast.}
    \label{fig:decoration-a}
\end{figure}

\begin{figure}
\includegraphics{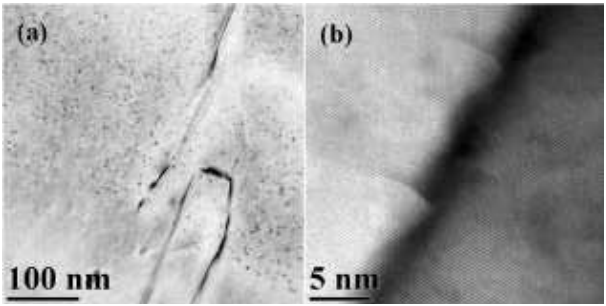}
\caption{\label{fig:defect} (a) Annihilation of two threading dislocations; (b) dislocation decorated by two distorted PIDs.}
\end{figure}

In our TEM study, we do not observe a significant reduction in the number of TD upon Mg-$\delta$-doping. The overall dislocation densities are in the order of 10$^{9}$ cm$^{-2}$ for edge dislocations, and 10$^{8}$ cm$^{-2}$ for screw- and mix-type dislocations, roughly the same as those estimated by AFM.
The weak-beam dark field images in Fig.~\ref{fig:dislocation}, are acquired at $\textit{g}$ = 10$\bar{1}$0 (a) and g = 0002 (b). According to the invisibility criterion, the screw dislocation with Burgers vector $b = [0002]$  and the edge dislocation with $\textit{b}$ = (1/3)[11$\bar{2}$0] are visible in Fig.~\ref{fig:dislocation}(a) and Fig.~\ref{fig:dislocation}(b), respectively.
Mixed dislocations with b = (1/3)[11$\bar{2}$3] can be identified in both images.
In all the investigated GaN:$\delta$-Mg samples, the TDs are not affected by the $\delta$-doping process. It is actually expected that $\delta$-doping is an effective way to reduce the threading dislocations density in GaN.\cite{CPC} However, the reduction highly depends on the growth mode of the $\delta$-layer, $\textit{e.g.}$ 2D or 3D nucleation can crop the TDs, whereas if the delta layers add in an atomic step-flow mode, their density will not be reduced.\cite{BDS} At our growth conditions, the Mg dopants tend to occupy the Ga sites without any nucleation, consequently the TDs are not reduced and for this reason the improved morphology of the GaN:$\delta$-Mg samples must be not TD-related.
TDs in GaN, acting as charging centers and gathering Ga vacancies, are likely to attract the Mg ions. In the TEM micrograph of a GaN:$\delta$-Mg sample above the solubility limit reported in Fig.~\ref{fig:decoration-a}(a), the PIDs formed in between two threading dislocations are displayed. These inversion domains have an average size of 25 nm, much greater than the one (5-10 nm) of the non-TD-related PIDs. A close inspection of the dislocation line reveals the bamboo-like void contrast reported in Fig.~\ref{fig:decoration-a}(b).
Similar structures have been reported for Mg-continuously-doped AlGaN, where TEM studies have shown that the threading dislocations present hollow cores - initiated by Mg segregation - promoting the precipitation of Mg.\cite{CBW,CWL} In GaN:$\delta$-Mg we expect an analogous mechanism: two dislocations may preferentially trap Mg adatoms during the growth interruption, the subsequent diffusion creates voids around the dislocation, and promotes the formation of large PIDs in the area around the faults.

The columnar structure formed by the complex of the two dislocations and the PIDs between them, terminates in a pit at the sample surface. This effect is due to the large volume of the inversion domains present inside the column, which reduce the overall growth rate, as an effect of the different growth rates of the two now coexisting polarities.\cite{web6,RM}
By comparing again Mg-continuously-doped GaN and AlGaN,\cite{RKN,CLJ} and the GaN:$\delta$-Mg structures under investigation, it becomes evident that in GaN:Mg uniform inversion domain columns are formed in between two dislocations, whereas in the $\delta$-doped samples larger PIDs are generated instead of an uniform inversion domain. We tend to explain this phenomenon assuming that the overlateral diffusion of Mg is enhanced in the considered area, but the vertical diffusion is still hampered by the GaN spacer, therefore, the uniformed inversion domains are unlikely to form.

The gathering of Mg ions by the threading dislocations may also result in an annihilation of faults. In Fig.~\ref{fig:defect}(a) a dipole loop at the interface between the GaN buffer and the GaN:$\delta$-Mg is reported. Here the dislocations are of \textit{c}-type, as determined by the invisiblility criterion. The density of PIDs is slightly reduced within the regions extending some tens of nanometers around the dislocations. Analogous annihilation loops have been reported for Si-$\delta$-GaN,\cite{CPC} and attributed to the pinning of the risers of the spiral step structures associated with screw dislocation. Similarly, during the growth interruptions of our GaN:$\delta$-Mg samples, the Mg ions may accumnulate at the end step-edges of screw dislocations. In the subsequent growth, then, the dislocation line is compelled to bend over the basal plane. In this way, two neighbouring screw dislocations with opposite Burgers vector can pair and annihilate, generating in this way a dislocation loop. It is worth to emphasize that the annihilation of dislocations by pinning is very occasional in GaN:$\delta$-Mg, and it does not affect the average dislocation density.

In a simpler case, the gathering of Mg ions by TDs merely causes a decoration of the dislocations, as shown in Fig.~\ref{fig:defect}(b). Here the two distorted PIDs are separated by a distance of 12 nm, corresponding to the thickness of the GaN spacer, making it likely that the PIDs are formed due to overlateral diffusion from the $\delta$-Mg layers to the dislocations.

\begin{figure}
\includegraphics{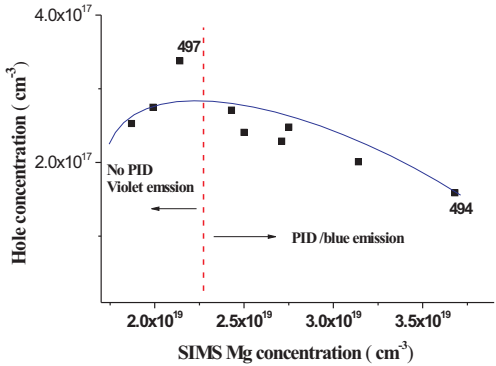}
\caption{\label{fig:HCC} GaN:$\delta$-Mg: free hole concentrations a function of the Mg content as determined by SIMS.}
\end{figure}

\subsection{Electrical and optical properties of GaN:$\delta$-Mg}

As previously mentioned, from the structural analysis we can infer the solubility limit of Mg into GaN:$\delta$-Mg at the growth conditions to be 2.14$\times$10$^{19}$ ions/cm$^{3}$. In the following, we will demonstrate that when this value is exceeded, the electrical and optical properties of the GaN:$\delta$-Mg epilayers are significantly affected.

\begin{figure}
    \centering
        \includegraphics{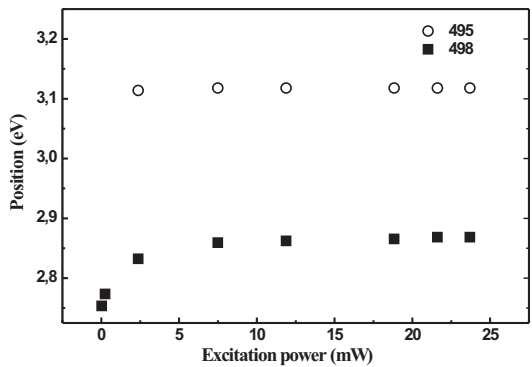}
    \caption{PL: peak energy of the VL and BL bands $\textit{vs.}$ excitation power for two GaN:$\delta$-Mg samples, respectively below (No.495) and above (No.498) the solubility limit of Mg.}
    \label{fig:fig_PL2}
\end{figure}
In Fig.~\ref{fig:HCC} the hole concentration at RT is reported as a function of the Mg content evaluated $\textit{via}$ SIMS. With increasing Mg effective incorporation, the hole concentration tends to saturate around 2.25$\times$10$^{19}$ Mg ions/cm$^{3}$ and then decreases, due to self-compensation upon phase separation.\cite{OBK} The highest hole concentration we achieved is 3.38$\times$10$^{17}$ cm$^{-3}$ for a Mg content of 2.14$\times$10$^{19}$ cm$^{-3}$.

\begin{figure}
\includegraphics{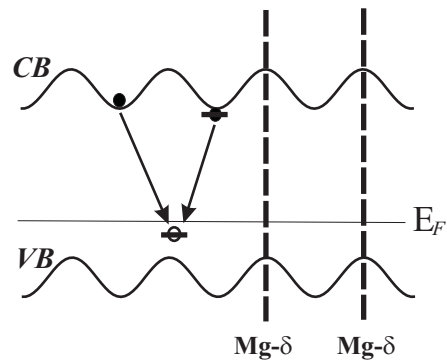}
\caption{\label{fig:Transition} GaN:$\delta$-Mg: schematic band diagram for the structure with periodic distribution of the Mg ions. In the GaN spacer between two consecutive Mg-$\delta$-doped layers, the decrement of the Mg concentration increases the Fermi level. The leveling of the Fermi levels across the whole heterostructure results in potential fluctuations that reduce the energy of free-to-bound transition.}
\end{figure}
\begin{figure}
    \centering
        \includegraphics{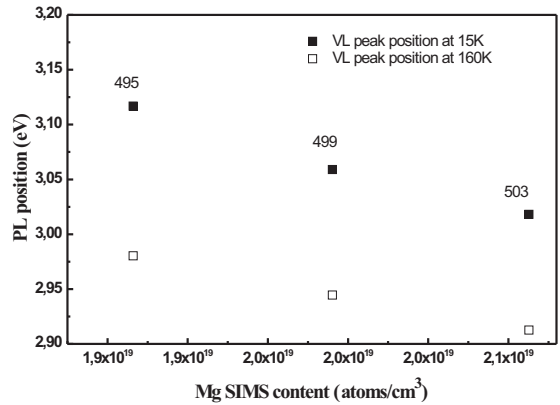}
    \caption{GaN:$\delta$-Mg: peak position of the VL band $\textit{vs.}$ Mg concentration at 15~K and 160~K, respectively.}
    \label{fig:PLa}
\end{figure}

The PL spectra of GaN:$\delta$-Mg reported in one of our previous works \cite{SWQ} present two main regimes as well: at concentrations of the Mg ions between 1.85 and 2.14$\times$10$^{19}$ cm$^{-3}$, a broad luminescence band at around 3.050~eV is formed, whereas for higher contents, a second band arises with a maximum in the blue spectral range at around 2.870~eV. The creation of these two bands is strictly limited to one of the two regimes, and coexistance could not be observed.
In order to gain insight into the nature of the radiative recombination mechanisms involved, we have studied the dependence on excitation power of the peak position for the violet luminescence (VL) and blue luminescence (BL) at 15K. As shown in Fig.~\ref{fig:fig_PL2}, the violet band presents a slight blue shift of about 20 meV, while the shift of the BL is up to 100~meV. We therefore attribute the VL and BL to free-to-bound transitions (Mg$^{0}$, e$^{-}$) and to deep donor-acceptor pair (DDAP) recombinations, respectively. The above mentioned recombination effects are similar to those reported for GaN:Mg. However, in the low doping regime, the transition energy in GaN:Mg is in the range of 3.1~-~3.25~eV, \cite{VSL,OPP,MSK} slightly higher than the one we observe in GaN:$\delta$-Mg. The different transition energy may be related to a dissimilar average Mg-concentration - usually not specified in literature - or to the modulated Mg distribution in Mg-$\delta$-doped GaN.
In the low doping regime, the diffusion of Mg into the GaN spacer layers is limited, the periodicity in the Mg ions concentration causes the Fermi level to oscillate with a period tuned with the $\delta$-layers occurrence. The leveling of the Fermi energy by the thermal equilibrium gives rise to the periodic potential fluctuations sketched in Fig.~\ref{fig:Transition}. The recombination between the electrons in proximity of the conduction band minimum and the deep Mg acceptors gives rise to a VL band with a reduced energy.
In Fig.~\ref{fig:PLa} the position of the VL is shown to shift to lower energies with increasing Mg concentration. Below the solubility limit, acceptor levels become shallower with increasing Mg concentration, leading to a blue-shift of the emission energy of the free-to-bound transition. The red-shift here indicates that the potential fluctuations are enhanced with the raising of the Mg concentration. As the Mg diffusion is limited in the low doping regime, the increasing of Mg content may amplify the potential fluctuation, generating an emission at an even lower energy.

\section{summary}

The structural properties of GaN:$\delta$-Mg grown on $c$-plane sapphire substrates by MOVPE have been investigated at the nanoscale and correlated with the electrical and optical response of these epilayers. Moreover, the characteristics of the $\delta$-doped samples have been compared with those of continuously Mg-doped GaN with equivalent Mg content, in order to gain insight into the effectiveness of $\delta$-doping for the required enhancement of the $p$-conductivity efficiency in nitride-based heterostructures. At the employed growth conditions, the solubility limit of Mg into GaN is reached for an average Mg concentration of 2.14 $\times$ 10$^{19}$ cm$^{-3}$. Above the solubility limit, Mg-rich PIDs are observed, with a density increasing with the average Mg concentration.
The surface of the GaN:$\delta$-Mg epilayers is found to be virtually free of extended defects, due to the hindering of the vertical diffusion of Mg caused by the GaN interlayers. By contrast, in continuously doped GaN the Mg dopants tend to diffuse to and accumulate on the surface, resulting in more extended defects (inversion domains). These facts may give a reason for the improved surface morphology of GaN:$\delta$-Mg as compared with that of GaN:Mg. We can not, however, confirm the reported reduction of threading dislocations as an effect of $\delta$-doping. The onset of phase separation in the layers affects the electrical and optical properties as well: $\textit{e.g.}$ self-compensation effects begin to influence the conductivity of the structures and the dominant recombination mechanisms take place in a different energy range.
We expect that, by tuning the growth parameters (including growth temperature, Cp$_{2}$Mg flow rate, interruption time and spacer layer thickness) during the fabrication of GaN:$\delta$-Mg, the solubility limit of Mg into GaN can be affected in a controlled way, allowing $\textit{e.g.}$ to enhance the incorporation of the magnetic ions, to hinder the self-compensation mechanisms and, therefore, to improve the concentration of active carriers in the layers. Furthermore, recent theoretical works postulate the onset of 3D interactions in $\delta$-doped nitride-based DMS, leading to an enhancement of the Curie temperature.\cite{FUKU} From the above considerations, it appears more and more clear that the $\delta$-doping technique, in combination with characterization methods at the nanoscale, can be further exploited and has the potential to promote the performance of functional future devices. 

\begin{acknowledgments}
The authors would like to thank T. Dietl for fruitful discussions, R. Jakiela for the SIMS measurements and G. Hesser for technical assistance. This work was financially supported by the Austrian Fonds zur Forderung der wissenschaftlichen Forschung - FWF (projects P17169-N08 and N107-NAN).

\end{acknowledgments}



\begin{thebibliography}{00000000}

\bibitem{TOM} J. W. Orton and C. T. Foxon, Rep. Prog. Phys. \textbf{61}, 1 (1998).

\bibitem{Dietl} T. Dietl, H. Ohno, F. Matsukura, J. Cibert, and D. Ferrand, Science \textbf{287}, 1019 (2000).

\bibitem{naka1}S. Nakamura, N. Iwasa, M. Senoh, and T. Mukai, Jpn. J. Appl. Phys. \textbf{31}, 1258 (1992).

\bibitem{amano} H. Amano, M. Kito, K. Hiramatsu, and I. Akasaki, Jpn. J. Appl. Phys. \textbf{28}, L2112 (1989).

\bibitem{naka}S. Nakamura, T. Mukai, M. Senoh, and N. Iwasa, Jpn. J. Appl. Phys. \textbf{31},L139 (1992).

\bibitem{NVM} J. Neugebauer, C. G. Van de Walle, Mater. Res. Soc. Symp. Proc. \textbf{449}, 509 (1996).

\bibitem{KSO} U. Kaufmann, P. Schlotter, H. Obloh, K. K\"{o}hler, and M. Maier, Phys. Rev. B \textbf{62}, 10867 (2002).

\bibitem{OBK} H. Obloh, K. H. Bachem, U. Kaufmann, M. Kunzer, M. Maier, A. Ramakrishnan, and P. Schlotter, J. Cryst. Growth \textbf{195}, 270 (1998).

\bibitem{NST} A. M. Nazmul, S. Sugahara, and M. Tanaka, Appl. Phys. Lett. \textbf{80}, 3210 (2002).

\bibitem{SKK} E. F. Schubert, J. M. Kuo, R. F. Kopf, H. S. Luftman, L. C. Hopkins, and N. J. Sauer, J. Appl. Phys. \textbf{67}, 1969 (1990).

\bibitem{CPC} O. Contreras, F. A.Ponce, J. Christen, A. Dadgar and A. Krost, Appl. Phys. Lett. \textbf{81}, 4712 (2002).

\bibitem{NKJ} M. L. Nakarmi, K. H. Kim, J. Li, J. Y. Lin, and H. X. Jiang, Appl. Phys. Lett. \textbf{82}, 3041 (2003).

\bibitem{WLN} H. Wang, J. Liu, N. Niu, G. Shen, and S. Zhang, J. Cryst. Growth \textbf{304}, 7 (2007).

\bibitem{web1} Z. Liliental-Weber, M. Benamara, W. Swider, J. Washburn, I. Grzegory, S. Porowski, D.J.H. Lambert, C.J. Eiting, and R.D. Dupuis, Appl. Phys. Lett. \textbf{75}, 4159 (1999).

\bibitem{web2} Z. Liliental-Weber, J. Jasinski, M. Benamara, L. Grzegory, S. Porowski, D. J. H. Lampert, C. J. Eiting, and R. D. Dupuis, Phys. Stat, Sol. (b) \textbf{228}, 345 (2001).

\bibitem{web3} Z. Liliental-Weber, M. Benamara, W. Swider, J. Washburn, I. Grzegory,
S. Porowski, R.D. Dupuis, and C.J. Eiting, Physica B \textbf{273-274},124 (1999).

\bibitem{SWQ} C. Simbrunner, M. Wegscheider, M. Quast, Tian Li, A. Navarro-Quezada, H. Sitter, and A. Bonanni, Appl. Phys. Lett.  \textbf{90}, 142108 (2007).

\bibitem{web5} Z. Lilitental-Weber, J. Jasinski, M. Benamara, I. Grzegory, S. Porowski, D. J. H. Lampert, C. J. Eiting, and R. D. Dupuis, Phys. Stat. Sol. (b) \textbf{228}, 345 (2001).

\bibitem{SIJAP} C. Simbrunner, H. Sitter, and A. Bonanni, J. Appl. Phys. \textbf{101}, 093501 (2007).

\bibitem{SIJCG} C. Simbrunner, K. Schmidegg, A. Kharchenko, J. Bethke, J. Woitek, K. Lischka, A. Bonanni, and H. Sitter, J. Cryst. Growth \textbf{298}, 243 (2007).

\bibitem{VBB} P. Venn$\acute{e}$gu$\grave{e}$s, M. Benaissa, B. Beaumont, E. Feltin, P. De Mierry, S.Dalmasso, M. Leroux, and P. Gibart, Appl. Phys. Lett. \textbf{77}, 880 (2000).

\bibitem{BVB} M. Benaissa, P. Venn$\acute{e}$gu$\grave{e}$s, B. Beaumont, P. Gibart, W. Saikaly, and A. Charai, Appl. Phys. Lett. \textbf{77}, 2115 (2000).

\bibitem{RFS} V. Ramachandran, R. M. Feenstra, W. L. Sarney, L. Salamanca-Riba, J. E.
Northrup, L. T. Romano, and D. W. Greve, Appl. Phys. Lett. \textbf{75}, 808 (1999).

\bibitem{RNP} L. T. Romano, J. E. Northrup, A. J. Ptak, and T. H. Myers, Appl. Phys.
Lett. \textbf{77}, 2479 (2000).

\bibitem{VBD} P. Venn$\acute{e}$gu$\grave{e}$s, M. Benaissa, S. Dalmasso, M. Leroux, E. Feltin, P. De Mierry, B. Beaumont, B. Damilano, N. Grandjean, and P. Gibart, Mater. Sci. Eng. B \textbf{93}, 224 (2002).

\bibitem{web4} Z. Lilitental-Weber, M. Benamara, J. Washburn, I. Grzegory, and S. Porowski, Phys. Rev. Lett. \textbf{83}, 2370 (1999).

\bibitem{PSD} A. Pretorius, M. Schowalter, N. Daneu, R. Kr$\ddot{o}$ger, A. Re$\check{c}$nik, and A. Rosenauer, Phys. Stat. Sol. (c)\textbf{3}, 1803 (2006).

\bibitem{LXZ} M. E. Lin, G. Xue, G. L. Zhou, J. E. Greene, and H. Morkoc, Appl. Phys. Lett. \textbf{63}, 932 (1993).

\bibitem{GBC} S. Guha, N. A. Bojarczuk, and F. Cardone, Appl. Phys. Lett. \textbf{71}, 1685 (1997).

\bibitem{YCO} Hongbo Yu, Deniz Caliskan, and Ekmel Ozbay, J. Appl. Phys. \textbf{100}, 033501 (2006).

\bibitem{MHB} T. Metzger, R. H\"{o}pler, E. Born, O. Ambacher, M. Stutzmann, R. St\"{o}mmer, M. Schuster, H. G\"{o}bel, S. Christiansen, M. Albrecht, and H.P. Strunk, Philos. Mag. A \textbf{77}, 1013 (1998).

\bibitem{BDS} J. Bai, M. Dudley, W. H. Sun, H. M. Wang, and M. Asif Khan, Appl. Phys. Lett. \textbf{88}, 051903 (2006).

\bibitem{CBW} D. Cherns, M. Q. Baines, Y. Q. Wang, R. Liu, F. A. Ponce, H. Amano, and I. Akasaki, phys. stat. sol. (b) \textbf{234}, 850 (2002).

\bibitem{CWL}D. Cherns, Y. Q. Wang, R. Liu and F. A. Ponce,Appl. Phys. Lett. \textbf{81}, 4541 (2002).

\bibitem{web6} Z. Lilitental-Weber, H. Sohn, N. Newman, and J. Washburn, J. Vac. Sci. Technol. B \textbf{13}, 1578 (1995).

\bibitem{RM} L. T. Romano and T. H. Myers, Appl. Phys. Lett. \textbf{71}, 3486 (1997).

\bibitem{RKN} L. T. Romano, M. Kneissl, J. E. Northrup, C. G. Van de Walle and D. W. Treat, Appl. Phys. Lett. \textbf{79}, 2734 {2001}

\bibitem{CLJ} H. K. Cho, J. Y. Lee, S. R. Jeon, and G. M. Yang, Appl. Phys. Lett. \textbf{79}, 3788 (2001).

\bibitem{WBS} Z. Liliental-Weber, M. Benamara, W. Swider, J. Washburn, I. Grzegory, S. Porowski, R. D. Dupuis, and C. J. Eiting, MRS Internet J. Nitride Semicond. Res. \textbf{5S1}, W9.7 (2000).

\bibitem{VSL} A. K. Viswanath, E. Shin, J. I. Lee, S. Yu, D. Kim, B. Kim, Y.Choi, and C.-H. Hong, J. Appl. Phys. \textbf{83}, 2272 (1998).

\bibitem{OPP} E. Oh, H. Park, and Y. Park, Appl. Phys. Lett. \textbf{72}, 70 (1998).

\bibitem{MSK} J. M. Myoung, K. H. Shim, C. Kim, O. Gluschenkov, K. Kim, S.Kim, D. A. Turnbull, and S. G. Bishop, Appl. Phys. Lett. \textbf{69}, 2722 (1996).

\bibitem{FUKU} T. Fukushima, K. Sato, H. Katayama-Yoshida, and P. H. Dederichs, Jpn. J. Appl. Phys. \textbf{45}, L416 (2006).

\end{thebibliography}

\end{document}